\documentclass[
    ,final            % use final for the camera ready runs
 ]
  {aipproc}

\layoutstyle{6x9}

\begin{document}

\title{Top polarization as a probe of new physics}

%<Replace this text with PACS numbers; choose from this list:
                %\texttt{http://www.aip.org/pacs/index.html}>}

\author{Rohini M Godbole}{
  address={Center for High Energy Physics, IISc, Bangalore, 560 012, India}
}

\author{Saurabh D Rindani}{
  address={Physical Research Laboratory, Navarangpura, Ahmedabad 380 009,
India}
}

\author{Kumar Rao}{
  address={Helsinki Institute of Physics, P.O. Box 64, Helsinki 00014, Finland}
%%  ,altaddress={<author1 address>} % additional visiting address
}

\author{Ritesh K Singh}{
  address={Institut f\"{u}r Theoretische Physik und Astronomie, Universit\"{a}t W\"{u}rzburg, 97074, Germany}
}

\begin{abstract}
 We investigate the effects of new physics scenarios containing a high mass vector resonance on top pair production at the LHC, using the polarization of the produced top. In particular we use kinematic distributions of the secondary lepton coming from top decay, which depends on top polarization, as it has been shown that the angular distribution of the decay lepton is insensitive to the anomalous $tbW$ vertex and hence is a pure probe of new physics in top quark production. Spin sensitive variables involving the decay lepton are used to probe top polarization. Some sensitivity is found for the new couplings of the top.
\end{abstract}

\maketitle

%%%%%%%%%%%%%%%%%%%%%%%%%%%%%%%%%%%%%%%%%%%%
%% MAINMATTER
%%%%%%%%%%%%%%%%%%%%%%%%%%%%%%%%%%%%%%%%%%%%

\section{Introduction}

  Though the mass of the top quark has been measured precisely, its coupling to other particles and production and decay schemes have not been probed in detail so far. The top's large mass $\sim 172$ GeV, and hence its large coupling to the Higgs and longitudinal component of the $W$ boson makes it a sensitive probe of the mechanism of electroweak symmerty breaking (ESB). Top quark properties are an important input to electroweak precision analysis. Alternative mechanisms of ESB and many new physics scenarios incorporate a special role for the top quark. It is thus important to determine if the couplings of the top follow those of the first two generations as in the Standard Model (SM) or different electroweak and/or strong couplings apply in this case.
  
  An important property of the top is that it decays before hadronization (with a lifetime $\tau _t \sim 10^{-25}$ which is an order of magnitude smaller than the hadronization lifetime of $1/\Lambda_{QCD}\sim 10^{-24}$). As a result the spin of the {\it bare} top is not diluted by hadronization, but is imprinted on characteristic angular distributions of its decay products, the $W$ boson and $b$ quark. For a review see \cite{bernreuther}. The degree of top polarization depends on its production process and thus can probe SM/BSM scenarios, since any couplings of the top to new particles can alter its degree of polarization. In this talk, we investigate the effects on top polarization, via decay lepton angular distributions, due to the presence of a high mass vector resonance, like the $Z_H$ in Littlest Higgs models.
  
\section{Top polarization and decoupled observables}

The polarization of the top is detected through the angular distribution of its decay products. %In the SM the $t$ decays almost exclusively to a $b$ and $W^+$ with a branching ratio of 0.998. The $W^+$ then decays to either $u\bar{d}$ (two jets) with a branching ratio of 2/3 or to a lepton and neutrino with a BR of 1/3 for each lepton.% 
Mass reconstruction is better for the  hadronic decay channel but it suffers from a large QCD background. The leptonic channel is cleaner but mass reconstruction is difficult due to missing energy. For a $t\bar{t}$ final state at a hadron collider the optimal detection channel is semileptonic with $t\to b l^{+} \nu_{l}$ and $\bar{t}\to b +2$ jets.

Top polarization can allow measurements of parameters of the model/new physics and can give more information on the production mechanism than just the cross section. It requires parity violation and thus measures left-right mixing and can probe $CP$ violation through dipole couplings. Even in parity conserving QCD, though the top is produced unpolarized (at tree level), there are different rates for opposite helicity and same helicity $t\bar{t}$ production, i.e, the spin of the $t$ is correlated with the $\bar{t}$.

Top spin can be measured by the angular decay distribution of the particle/jet $f$ in the rest frame of the top. %,which in the top rest frame has the generic form \cite{bernreuther}
%\begin{equation}
% \frac{1}{\Gamma_f}\frac{d\Gamma_f}{d \cos \theta_f}=\frac{1}{2}(1+P_t \kappa_f \cos %\theta_f) \label {spinanalyser}
%\end{equation}
%where $\theta_f$ is the angle between the direction of $f$ in the top rest frame and it's polarization vector. $P_t$ is the degree of top polarization for an ensemble of top quarks and $\kappa_f$ is the $t$ ``spin analysing power'' of $f$. At tree level, $\kappa_{l^+}=-\kappa_d =1$, whereas $\kappa_b=\kappa_{W^+}=-0.39$. Thus the charged lepton or $d$ quark has the best spin analysing power, with the probablity of the $l^{+}/d$ being emitted in the direction of the $t$ spin being maximal while being zero in the direction opposite to the $t$ spin. 
For hadronic $t\bar{t}$ production, spin correlations between the decay leptons from the $t$ and $\bar{t}$ have been extensively studied \cite{bernreuther,spincorr}. These spin correlations measure the asymmetry between the production of like and unlike helicity pairs of $t\bar{t}$ which can probe new physics in top pair production. However, this requires the reconstruction of the $t$ and $\bar{t}$ rest frames, which is difficult, if not impossible, at the LHC. Here we investigate if {\it single top polarization} can be a qualitative or quantitative probe of new physics and can provide better statistics for top pair production at the LHC.

New physics can also appear in the $tbW$ decay vertex, apart from that in top production, leading to changed decay width and distributions for the $W^+$ and $l^+$. The model independent form factor for the $tbW$ vertex can be parametrised as
\begin{equation}
 \Gamma^\mu =\frac{-ig}{\sqrt{2}}\left[\gamma^{\mu}(f_{1L} P_{L}+f_{1R}P_{R})-\frac{i \sigma^{\mu \nu}}{m_W}(p_t -p_b)_{\nu}(f_{2L}P_{L}+f_{2R}P_{R})\right] \label{anomaloustbW}
\end{equation}
where for the SM $f_{1L}=1$ and the anomalous couplings $f_{1R}=f_{2L}=f_{2R}=0$. 
The simultaneous presence of new physics in top production and decay can complicate the analysis making it difficult to isolate new couplings of the top. However it has been proved that {\it angular distributions} of charged leptons/$d$ quarks from top decay are {\it not affected by the anomalous $tbW^{+}$ vertex}, see \cite{saurabh} and references therein. This has been shown very generally for a $2\to n$ process and assumes the narrow width approximation (NWA) for the top and neglects terms quadratic in the anomalous couplings in (\ref{anomaloustbW}), assuming new physics couplings to be small. This implies that charged lepton angular distributions, a {\it decoupled observable}, are more accurate probes of top polarization, and thus to new physics in top production alone. In contrast, the energy distributions of the $l^+$ or the angular distributions of the $b$ and $W$ are ``contaminated'' by the anomalous $tbW$ vertex.
%%\subsection{<A subsection>}
%%\paragraph{<A subsubsubsection>}
%%%%%%%%%%%%%%%%%%%%%%%%%%%%%%%%%%%%%%%%%%%%
%% Sample figure:
%%
%% The option [height=...] scales the picture to the given height,
%% without it it would be printed at its nominal size
%%%%%%%%%%%%%%%%%%%%%%%%%%%%%%%%%%%%%%%%%%%%
\begin{figure}
\includegraphics[height=.2\textheight,width=.35\textheight]{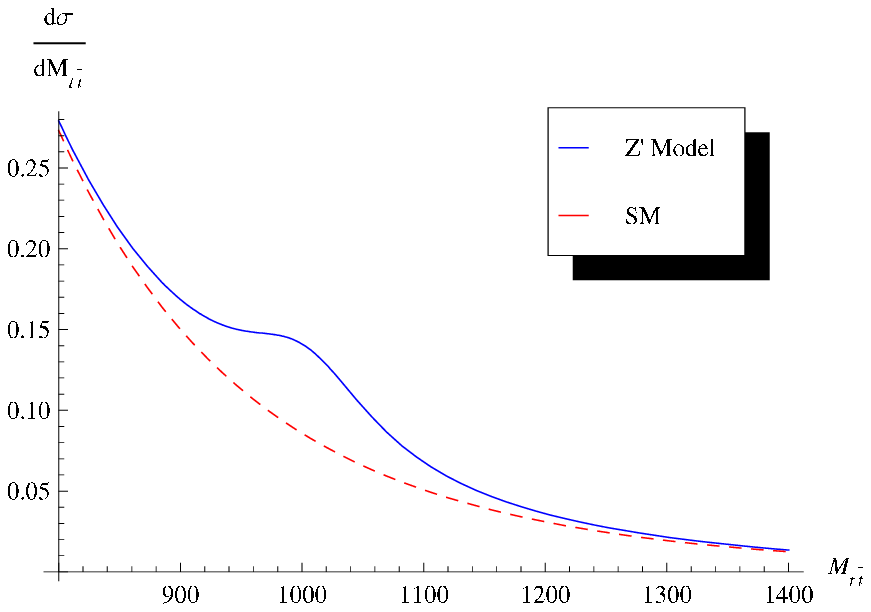}
\includegraphics[height=.2\textheight,width=.35\textheight]{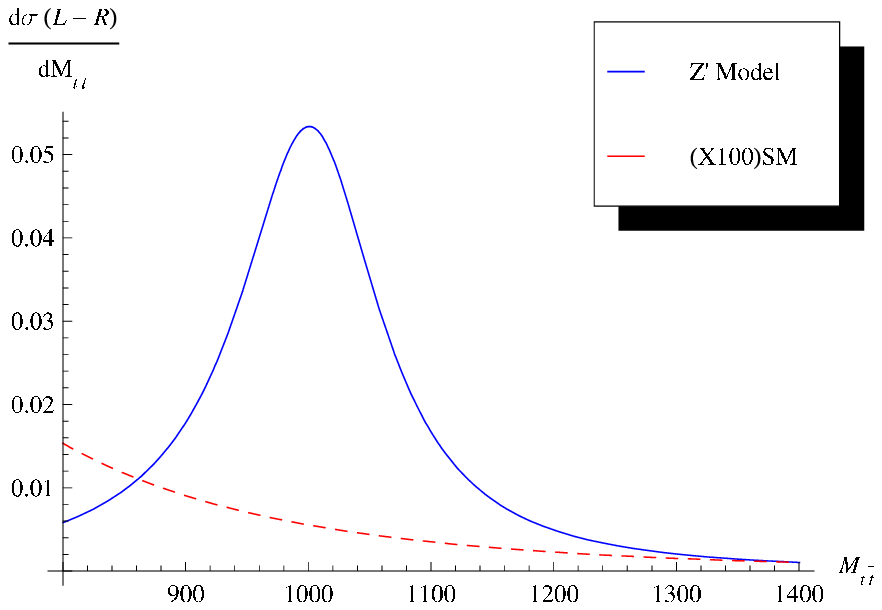}
  \caption{The $t\bar{t}$ invariant mass distribution in GeV for top pair production at the LHC for the total (left) and polarized (right) cross sections with $M_{Z_H}=1000$ GeV and $\cot \theta =2$.}
\end{figure}
%\begin{figure}
% \includegraphics[height=.2\textheight,width=.35\textheight]{Pt.eps}
%\end{figure}
Using the spin density matrix formalism for top production and decay (which retains the spin information of the decaying quark) and the NWA, the amplitude squared can be factored into production and decay parts as
\begin{equation}
 |\mathcal{M}|^2 = \frac{\pi \delta (p_t^2 -m_t^2)}{m_t \Gamma_t} \sum _{\lambda , \lambda^{'}} \rho (\lambda , \lambda^{'}) \Gamma (\lambda , \lambda^{'})
\end{equation}
where $\rho (\lambda , \lambda^{'})$ and $\Gamma (\lambda , \lambda^{'})$ are the $2 \times 2$ production and decay spin density matrices. After phase space integration of $\rho (\lambda , \lambda^{'})$ the resulting top production density matrix $\sigma (\lambda , \lambda^{'})$ can be used to define the polarization density matrix
\begin{eqnarray}
P_t = \frac{1}{2}\left(
\begin{tabular}{cc}
$1+\eta_3$ & $\eta_1 - i\eta_2$\\ 
$\eta_1 + i\eta_2$ & $1-\eta_3$
\end{tabular} \right),
%\label{poldm}
\end{eqnarray}
where the top longitudinal polarization is $\eta_3=(\sigma(+,+)-\sigma(-,-))/\sigma_{\rm{tot}}$. The transverse polarization in the production plane is $\eta_1=(\sigma(+,-)+\sigma(-,+))/\sigma_{\rm{tot}}$ while the transverse polarization perpendicular to the production plane is $i \eta_2=(\sigma(+,-)-\sigma(-,+))/\sigma_{\rm{tot}}$. The $\eta$'s can be calculated from the decay lepton angular distributions by a suitable combination of polar and azimuthal angular asymmetries, for details see \cite{saurabh}. This requires partial or complete reconstruction of the top momentum in the lab frame. Alternatively, the azimuthal distribution of the lepton in the lab frame can be used to probe top polarization.

\section{Top pair production at the LHC in $Z'$ models}
$t \bar{t}$ production at the LHC proceeds therough $gg,\,q\bar{q} \to t \bar{t}$. In models with an extra heavy gauge boson, $Z'$, top pair production receives an extra contribution in the $s$ channel apart from the $\gamma,Z$ and gluon. For concreteness we choose the Littlest Higgs Model where the left handed couplings to quarks of the resonance $Z_H$ can be parametrized as $(v_f,\,a_f)=\pm (g \cot \theta /4,\, g \cot \theta /4)$ for $T_3=\pm 1/2$, where $g$ is the electroweak coupling and $\theta$ is a mixing angle in the model \cite{tao}. The $Z_H$ mass and coupling $\cot \theta$ completely specify $t \bar{t}$ production and decay dynamics. Figure. 1 shows the effects of the $Z_H$ resonance on the total cross section and the difference between the cross sections for the left and right handed top's. Understandably, the resonace is sharper for the polarized cross section because of the absence of the large QCD background.
\begin{figure}
\includegraphics[height=.22\textheight,width=.30\textheight]{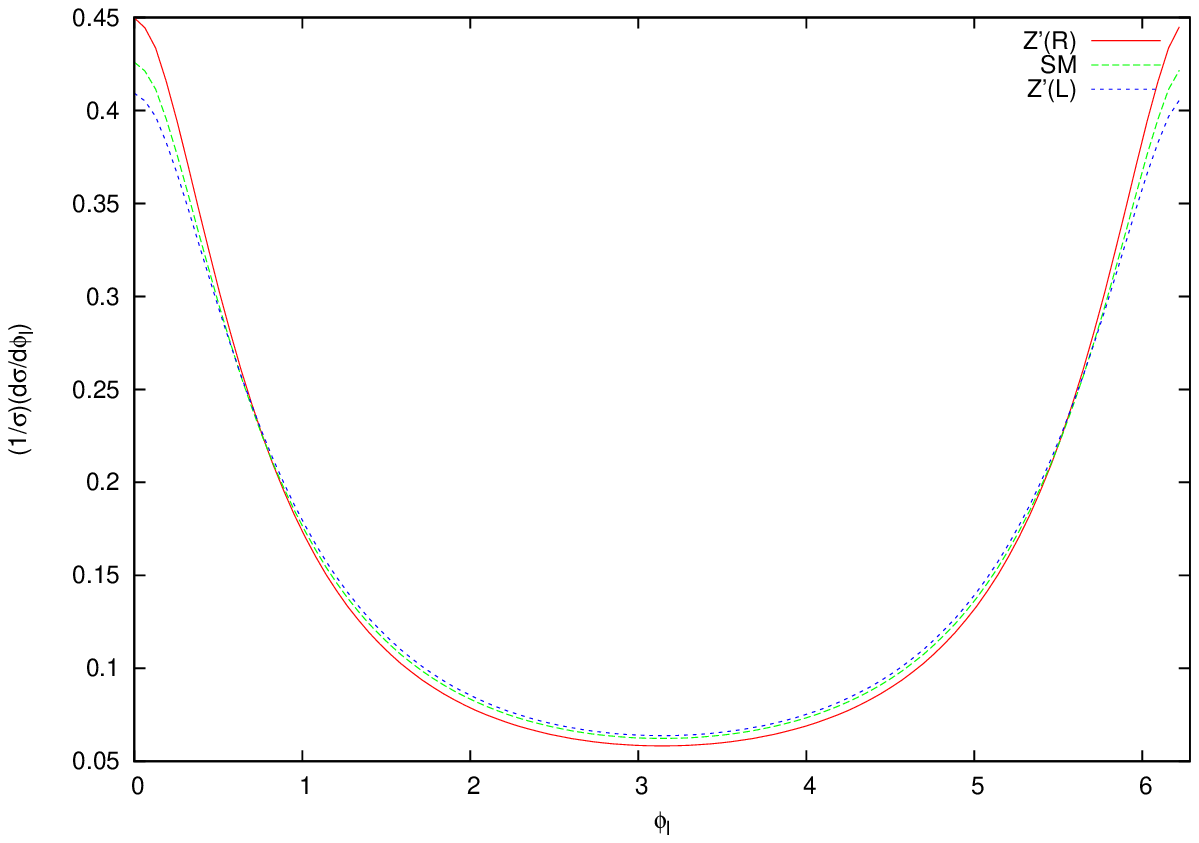}
\includegraphics[height=.22\textheight,width=.32\textheight]{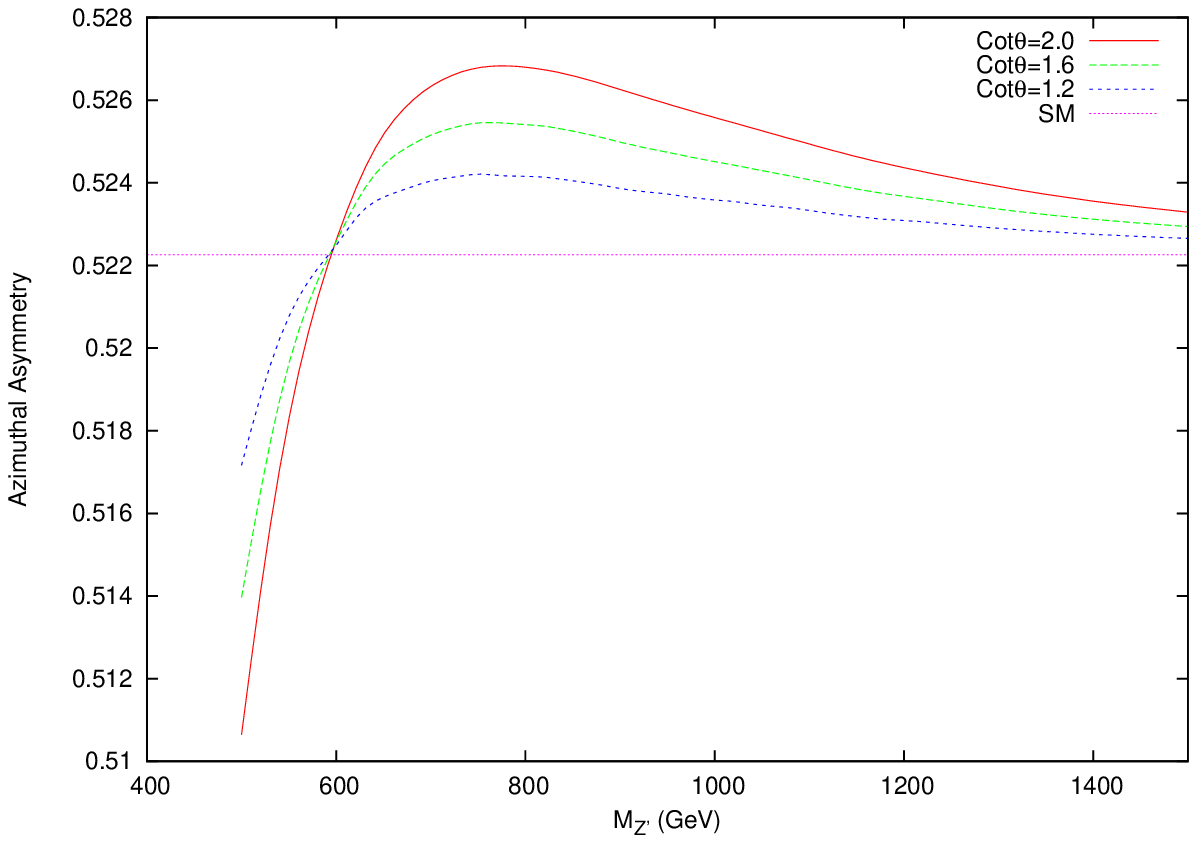}
\caption{The $\phi_l$ distribution for the SM and Littlest Higgs model for right and left handed top-$Z_{H}$ couplings (left) and the azimuthal asymmetry $A$ as a function of $M_{Z_{H}}$ (right).}
\end{figure}
We now consider angular distributions of the lepton in the lab frame, since there are easily measured experimentally at the LHC (considering inclusive decay of the $\bar{t}$). They are a pure probe of top polarization, since they are independent of any possible new physics in top decay. However, the polar angle, measuring the forward backward asymmetry with respect to boost direction in the parton cm frame, is only mildly sensitive to the magnitude of top polarization and almost insensitive to the sign of polarization. This leaves the lepton azimuthal angle $\phi_l$ measured with respect to the $t \bar{t}$ production plane. Figure. 2 shows the normalized $\phi_l$ distribution for the SM and the Littlest Higgs for $M_{Z_{H}}=500$ GeV and $\cot \theta =2$. The peak at $\phi_l=0, 2 \pi$ is sensitive to the magnitude of top polarization $\eta_3$ and can distinguish between right and left handed tops ($\eta_2=0$ in this case resulting in the $\phi_l$ distribution being symmetric around $\pi$). We can thus construct a normalized azimuthal asymmetry sensitive to top polarization for comparision with the SM case, by taking the difference between cross sections in the (1,4) and (2,3) quadrants: $A\equiv (\sigma(1,4)-\sigma(2,3))/(\sigma(1,4)+\sigma(2,3))$. A plot of the azimuthal asymmetry for different $\cot \theta$ is shown in Fig. 2 as function of $Z_H$ mass. $A$ reflects the degree of top polarization, saturating to the SM value for large $M_{Z_{H}}$ and equalling it around 600 GeV. We find a large sensitivity for the variable $O\equiv A -A_{SM}$, the difference between the azimuthal asymmetries in the Littlest Higgs and SM. %reflects well the degree of top polarization. %Lepton angular distributions can be used to get quantitative information about top polarization, i.e, the $\eta_i$, by reconstructing the top momentum in the lab frame, as shown in \cite{saurabh}. 
Work is currently underway to study spin sensitive variables to qualitatively and quantitatively probe top polarization.% and new top couplings.
%%%%%%%%%%%%%%%%%%%%%%%%%%%%%%%%%%%%%%%%%%%%%%%%
%% BACKMATTER
%%%%%%%%%%%%%%%%%%%%%%%%%%%%%%%%%%%%%%%%%%%%%%%%

\begin{theacknowledgments}
 We  acknowledge support from the Indo-French project IFCPAR No. 3004-2. RMG acknowledges DST, India for support under Grant No. SR/S2/JCB-64/2007 and KR thanks the Academy of Finland for partial support under project No. 115032. The work of R.K.S. is supported by the German BMBF under contract 05HT6WWA.
\end{theacknowledgments}


\bibliographystyle{aipproc}   % if natbib is available

%\bibliography{sample}

\begin{thebibliography}{9}

\bibitem{bernreuther}
 W.~Bernreuther,
  %``Top quark physics at the LHC,''
  J.\ Phys.\ G {\bf 35}, 083001 (2008)
  [arXiv:0805.1333 [hep-ph]]; M. Beneke {\t et al.} 2000 [arXiv:hep-ph/0507207]; S.~Willenbrock,
  %``The Standard model and the top quark,''
  arXiv:hep-ph/0211067.
  %%CITATION = HEP-PH/0211067;%%
 
  %%CITATION = JPHGB,G35,083001;%%

\bibitem{spincorr} G. Mahlon and S.J. Parke, Phys. Rev. D {\bf 53}, 4886 (1996) [arXiv:hep-ph/9512264]; Phys. Lett. B {\bf 411}, 173 (1997) [arXiv:hep-ph/9706304]; %%T. Stelzer and S. Willenbrock, Phys. Lett. B {\bf 374}, 169 (1996) [arXiv:hep-ph/9512292]; W. Bernreuther, A. Brandenberg, Z.G. Si and P. Uwer, Phys. Rev. Lett. {\bf 87}, 242002 (2001) [arXiv:hep-ph/0107086]; Nucl. Phys. B {\bf 690}, 81 (2004) [arXiv:hep-ph/0403035].

\bibitem{saurabh}
  R.~M.~Godbole, S.~D.~Rindani and R.~K.~Singh,
  %``Lepton distribution as a probe of new physics in production and decay of
  %the t quark and its polarization,''
  JHEP {\bf 0612}, 021 (2006)
  [arXiv:hep-ph/0605100].
  %%CITATION = JHEPA,0612,021;%%

\bibitem{tao} T.~Han, H.~E.~Logan, B.~McElrath and L.~T.~Wang,
  %``Phenomenology of the little Higgs model,''
  Phys.\ Rev.\  D {\bf 67}, 095004 (2003)
  [arXiv:hep-ph/0301040].
  %%CITATION = PHRVA,D67,095004;%%



\end{thebibliography}

\IfFileExists{\jobname.bbl}{}
 {\typeout{}
  \typeout{******************************************}
  \typeout{** Please run "bibtex \jobname" to optain}
  \typeout{** the bibliography and then re-run LaTeX}
  \typeout{** twice to fix the references!}
  \typeout{******************************************}
  \typeout{}
 }

\end{document}